\UseRawInputEncoding 

\documentclass{nature-pre}
\usepackage{amsmath}
\usepackage{amssymb}
\usepackage{graphicx}

\usepackage{xcolor}

\usepackage{notes2bib}

\usepackage{underscore}
\usepackage[english]{babel}

\usepackage{multirow}

\usepackage{tabularx} 

\usepackage{comment}

\graphicspath{./}



\title{Quantum sensor in a single layer van der Waals material}

\author{Rohit Babar$^{1,\dagger}$, Gergely Barcza$^{2,\dagger}$, Anton Pershin$^{2,\dagger}$,  Hyoju Park$^3$,  Oscar Bulancea Lindvall$^{1}$, Gerg{\H o} Thiering$^{2}$,  \"{O}rs Legeza$^{2,4,5}$,  Jamie H. Warner,$^{3}$  Igor A.  Abrikosov$^{1}$, Adam Gali$^{2,6,*}$,  and Viktor Iv\'{a}dy$^{1,2,7,*}$}

\begin{document}
\maketitle

\begin{affiliations}
\item   {Department of Physics, Chemistry and Biology, Link\"oping University, SE-581 83 Link\"oping, Sweden}
\item  {Wigner Research Centre for Physics, PO Box 49, H-1525, Budapest, Hungary}
\item {Walker Department of Mechanical
Engineering and Materials Graduate Program, Texas
Materials Institute, The University of Texas at Austin, Austin, Texas 78712, United States}
\item  {Fachbereich Physik, Philipps-Universit\"at Marburg, 35032 Marburg, Germany}
\item  {Institute for Advanced Study, Technical University of Munich, Lichtenbergstrasse 2a, 85748 Garching, Germany}
\item   {Department of Atomic Physics, Budapest University of Technology and Economics, Budafoki \'{u}t 8., H-1111, Budapest, Hungary}
\item   {Max-Planck-Institut f\"{u}r Physik komplexer Systeme, N\"{o}thnitzer Street 38, D-01187 Dresden, Germany}
\item[$\dagger$] Contributed equally.
\item[*] email:gali.adam@wigner.hu; viktor.ivady@liu.se
\end{affiliations}

\date{\today}

\newpage

\begin{abstract}

Point defect qubits in semiconductors have demonstrated their outstanding high spatial resolution sensing capabilities of broad multidisciplinary interest. Two-dimensional (2D) semiconductors hosting such sensors have recently opened up new horizons for sensing in the subnanometer scales in 2D heterostructures. However, controlled creation of quantum sensor in a single layer 2D materials with high sensitivity has been elusive so far. Here, we report on a novel 2D quantum sensor, the VB2 centre in hexagonal boron nitride (hBN), with superior sensing capabilities. The centre’s inherently low symmetry configuration gives rise to unique electronic and spin properties that implement a qubit in a 2D material with unprecedented sensitivity. The qubit is decoupled from its dense spin environment at low magnetic fields that gives rise to the reduction of the spin resonance linewidth and elongation of the coherence time. The VB2 centre is also equipped with a classical memory that can be utilized in storing population information. Using scanning transmission electron microscopy imaging, we confirm the presence of the point defect structure in free standing monolayer hBN created by electron beam irradiation. Our results provide a new material solution towards atomic-scale sensing in low dimensions.

\end{abstract}

\newpage


\section*{Introduction}

In the new era of quantum sensing, point defect qubits play a crucial role in revolutionizing measurements in material science, biology, and medicine~\cite{degen_quantum_2017}. In particular, the NV center in diamond~\cite{DohertyNVreview} has provided the means to detect magnetic field~\cite{taylor_high-sensitivity_2008}, electric field~\cite{michl_robust_2019}, strain~\cite{ovartchaiyapong_dynamic_2014}, and temperature~\cite{Kucsko2013} with high spatial resolution and high sensitivity. Novel magnetic resonance protocols~\cite{devience_nanoscale_2015,bucher_quantum_2019} with unlimited frequency resolution~\cite{schmitt_submillihertz_2017,boss_quantum_2017} have opened new horizons for magnetic resonance in the few-spin limit that provides hitherto inaccessible information on the structure and functioning of molecules and proteins~\cite{lovchinsky_nuclear_2016}.  On the other hand, the NV centre is an inherently bulk system, that implies severe challenges when near surface functioning is demanded~\cite{bluvstein_extending_2019,dwyer_probing_2021}. Wide-band gap van der Waals semiconductors with mature exfoliation possibilities, such as the hexagonal boron nitride (hBN), hosting applicable, optically addressable point defect qubits on the surface and even in a single layer may be advantageous in various sensing applications~\cite{tetienne_quantum_2021}. 

Indeed, hBN has received considerable attention  in the domain of point defect based quantum information processing and quantum sensing applications~\cite{tran_quantum_2016,froch_coupling_2020}. There have been several bright single photon emitters reported with photon energy ranging from near infrared to ultraviolet~\cite{tran_quantum_2016,hayee_revealing_2020,sajid_single-photon_2020}. The photoluminescence signal of these centres often exhibits giant shift upon applied strain and electric field~\cite{li_giant_2020}. Recent optically detected magnetic resonance (ODMR) and magnetic field dependent photoluminescence studies have demonstrated optically addressable point defect spins in hBN that provides additional degrees of freedom for sensing~\cite{exarhos_magnetic-field-dependent_2019,gottscholl_initialization_2020,chejanovsky_single_2019,stern_room-temperature_2021}. One of the centres has already been assigned to the negatively charged boron vacancy (V$_{\text{B}}$) centre~\cite{gottscholl_initialization_2020,ivady_ab_2020,sajid_edge_2020,Reimers2020}.  Very recently, the V$_{\text{B}}$ centre has been coherently manipulated~\cite{gottscholl_room_2021,liu_rabi_2021} and proposed as a high spatial resolution sensor for temperature, deformation, and magnetic field~\cite{gottscholl_spin_2021}. The sensitivity of this novel sensors, however, most often falls behind the sensitivity of the NV centre in diamond~\cite{gottscholl_sub-nanoscale_2021}. The fundamental reasons behind the relatively low sensitivity of the V$_{\text{B}}$ centre are the low photon emission rate due to the first-order forbidden optical transition~\cite{ivady_ab_2020} and the noisy environment due to the 100\% natural abundance of paramagnetic isotopes in hBN.  In order to significantly enhance sensitivity, novel qubits are required that enable fast and high contrast read-out and decouple the spin state from the spin bath of the host material.   

Here we experimentally and computationally investigate a point defect complex created by the fusion of two boron vacancies in the neutral charge state. We demonstrate by annular dark-field scanning transmission electron microscopy (ADF-STEM) imaging that the stable configuration of the defect, named as VB2 center, can be created in single layer free-standing hBN samples. We report on nearly degenerate decoupled singlet-triplet pair states throughout the spectrum of the defect that enable long-lived storage of population information in the singlet manifold and give rise to electric dipole allowed optical transitions for reading out the spin states.  Furthermore, we demonstrate that the transverse zero-field splitting interaction of the ground state gives rise to efficient coherence protection mechanisms. As a result, the electron spin resonance linewidth is reduced by a factor of $4.2$ and the inhomogeneous coherence time is elongated by a factor of $7.5$ at zero magnetic field that directly enhances the sensitive of spin resonance based sensing applications. The reported favourable properties of the VB2 defect may be utilized in novel quantum sensing applications with an expected gain in sensitivity, exceeding two orders of magnitude relative to the boron vacancy in hBN.

\section*{Main}
\label{sec:res}

\subsection{Structure and stability}
\label{sec:struct}

\begin{figure}[h!]
\begin{center}
\includegraphics[width=0.8\columnwidth]{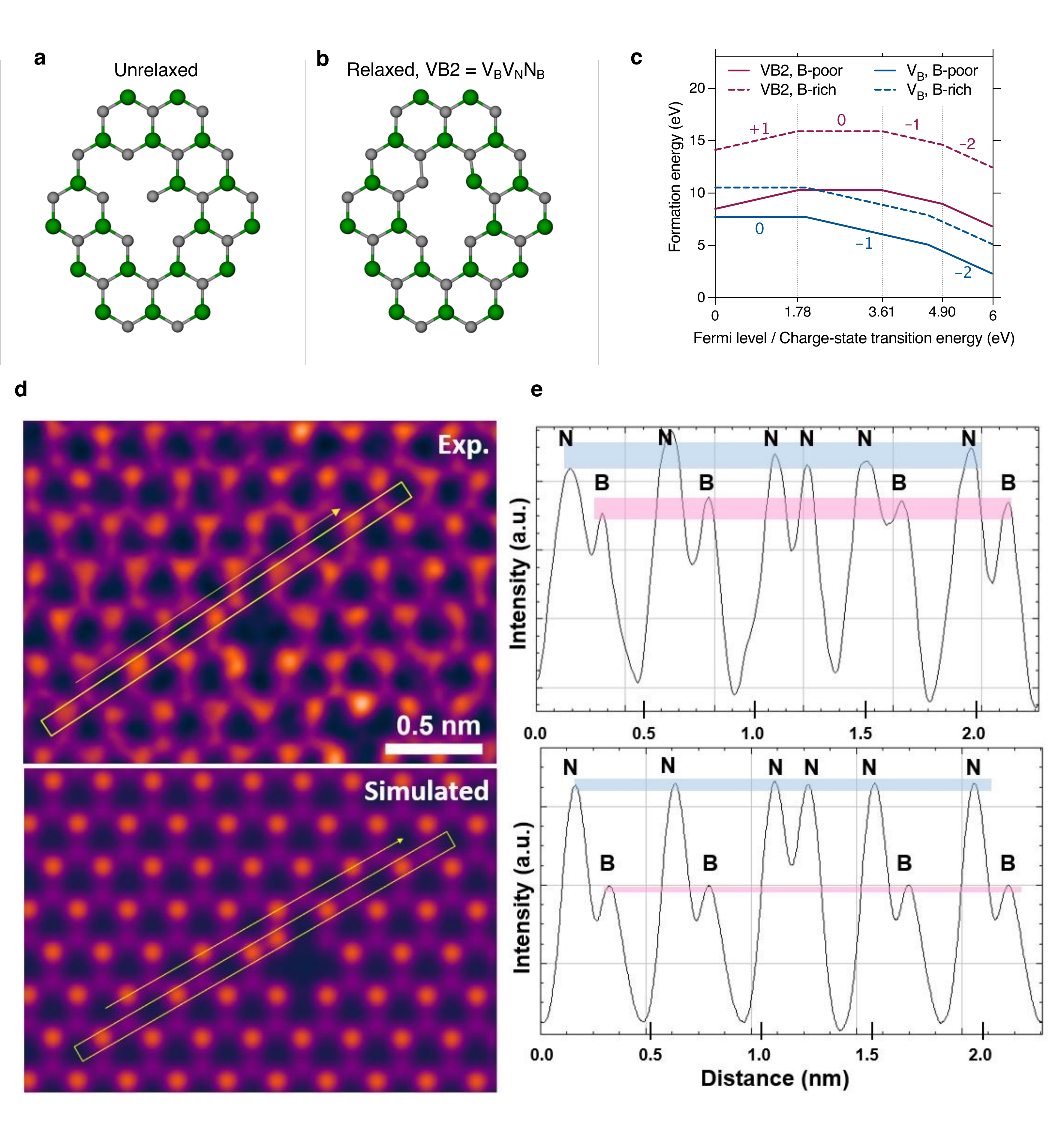}
\caption{ {\bf Structure and stability of the VB2 centre. }  {\bf a},  Unrelaxed structure of two adjacent boron vacancies. {\bf b}, The energetically most favourable VB2 configuration where the unstable nitrogen atom of two dangling bonds has jumped into the upper boron vacancy site forming a V$_{\text{B}}$V$_{\text{N}}$N$_{\text{B}}$ complex. {\bf c},  Formation energies of VB2 (magenta) and V$_B$ (light blue) defects in h-BN as a function of Fermi level under B-poor (solid line) and B-rich (dashed line) growth conditions. Colour coded numbers represent the charge states of the corresponding defect, while vertical dotted gray lines indicate the charge state transition levels of the VB2 defect. {\bf d}, A false-coloured ADF-STEM image of the VB2 configuration in single layer free standing hBN sample. Top and bottom panels show experiment and simulation of the VB2 configuration. {\bf e}, Intensity profile across the bright atoms along the yellow dashed box in {\bf d} that reveals an anti-site nitrogen next to the vacancy sites.  
\label{fig:fig1}  }
\end{center}
\end{figure}

A pair of adjacent in-plane boron vacancies results in a metastable configuration where the nitrogen atom between the two vacancies possess two dangling bonds and binds only to a single boron atom, see Fig.~\ref{fig:fig1}a. The loosely bonded nitrogen atom between the boron vacancies migrates into one of the vacancy sites and forms a new point defect complex consisting of a boron vacancy, a nitrogen vacancy, and a nitrogen anti-site (V$_{\text{B}}$V$_{\text{N}}$N$_{\text{B}}$), see Fig~\ref{fig:fig1}b. The defect complex is referred hereinafter as the VB2 configuration. Due to the appearance of strong nitrogen-nitrogen bonds in the VB2 configuration, the formation of the complex is highly favourable. Indeed, the binding energy of the vacancy pair is found to be as large as $5.14$~eV~\cite{Strand_2019}. 

The formation energy curves of the VB2 and the V$_\text{B}$ defects as a function of the Fermi energy are depicted in Fig.~\ref{fig:fig1}c. The figure demonstrates that the VB2 configuration exhibits four stable charge states and three charge-state transition levels, $(+1/0)$, $(0/\!\! - \!\! 1)$, and $(-1/\!\! - \!\! 2)$ at $1.78$~eV, $3.61$~eV, and $4.90$~eV, respectively, in the band gap of hBN. The neutral charge state is stable in the middle of the band gap and photo-stable up to $3.61$~eV photon energy, which is the threshold for exciting an electron from the conduction band to a defect ionisation state, see Fig.~\ref{fig:fig1}c.  

The VB2 configuration is directly observed using ADF-STEM imaging of a single layer free standing hBN sample, see Fig.~\ref{fig:fig1}d.  The image of the VB2 configuration resembles the boron vacancy-nitrogen vacancy complex, however, the intensity profile across the bright atoms shows an excellent match to the VB2 complex with an N atom contrast from one of the B atom sites, by comparing to the multi-slice simulated STEM image in Fig.~\ref{fig:fig1}e. We note that the appearance of the VB2 configuration is a direct consequence of the STEM measurement itself. In our measurements, a nearly defect free hBN layer is suspended on a metallic grid. As the electron beam scans through the sample defects, more likely boron vacancies are created~\cite{park_atomically_2021}. Prior work showed that electron beam irradiation of monolayer hBN causes ejection of B and N atoms around point defects.  A range of different small point defects result, and in some cases the VB2 defect is produced. We note also that high positioning accuracy may be achieved in defect fabrication in STEM by using focused electron beam pulses at specific points instead of scanning the surface~\cite{park_atomically_2021}. The VB2 configuration may also be realized in bulk hBN by either electron or neutron irradiation and subsequent annealing at $1000$~K. For further details on fabrication, see Supplementary Note 2 .

\subsection{Excited state properties}
\label{sec:opt}

The in-plane and perpendicular to plane dangling bonds of the vacancy sites of the VB2 configuration give rise to numerous localized defect states, most of which can be found in the band gap of hBN. A comprehensive analysis of the single particle electronic structure can be found Supplementary Note 3. As a consequence of the complicated electronic structure,  several excited states can be found below the ionisation threshold that results in diverse optical excitation and decay pathways. 

\begin{figure}[h!]
\begin{center}
\includegraphics[width=1.0\columnwidth]{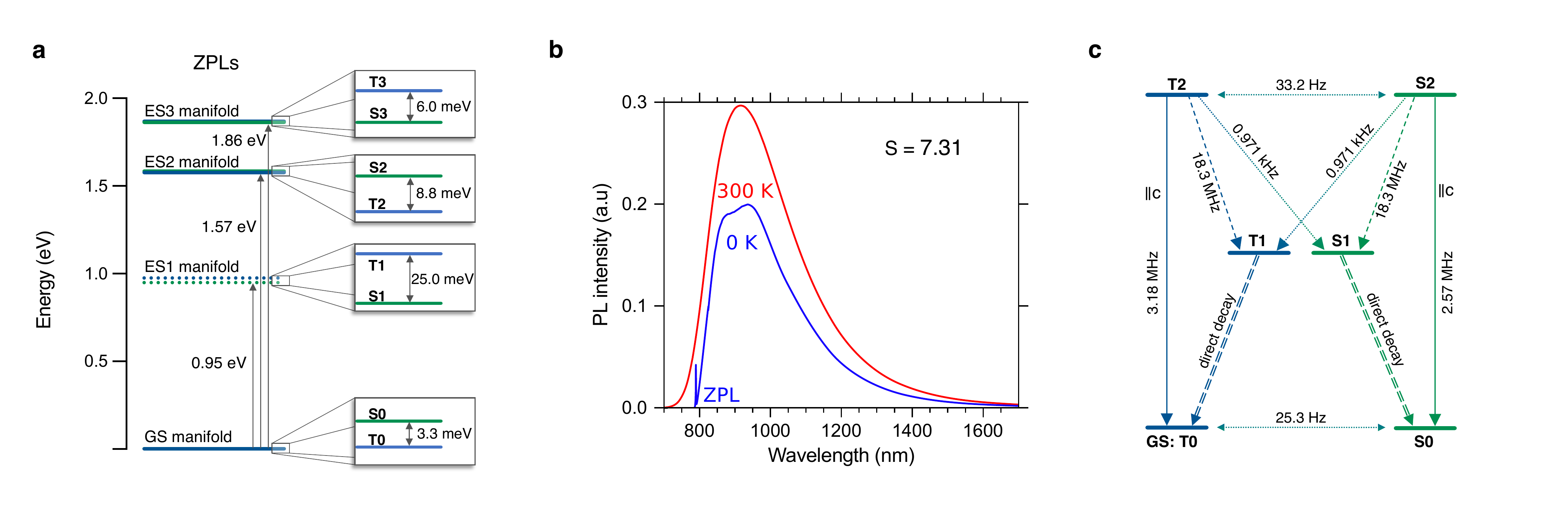}
\caption{ {\bf  Optical properties, initialisation, and non-destructive readout of VB2.  } {\bf a}, Energy level structure of the VB2 defect exhibiting quasi-degenerate singlet-triplet pairs throughout the spectrum. {\bf b},  PL emission spectra of the VB2 defect at room temperature. The emission maximum is red-shifted by $60$~nm compared to the PL emission peak of the boron vacancy center.  {\bf c}, Schematic diagram of the possible radiative and non-radiative relaxation pathways. Solid arrows indicate optical transition with parallel to $c$-axis polarization, dashed arrows indicate spin-conserving non-radiative relaxation, dashed double-line arrows indicate direct spin-conserving relaxation to the ground state, and dotted arrows indicate spin non-conserving non-radiative transitions through inter-system crossings. 
\label{fig:fig2}  }
\end{center}
\end{figure}

The energy level spectrum including the eight lowest lying many-particle energy levels obtained at the corresponding relaxed geometries is depicted in Fig.~\ref{fig:fig2}a.  The spectrum exhibits a peculiar feature not seen in point defect qubits before: it is built up from nearly degenerate pairs of singlet and triplet states, referred as excited manifolds hereinafter. The reason behind the observed quasi-degenerate singlet-triplet pairs can be understood by examining the ground state triplet. Due to the defect's low symmetry, the observed high-spin ground state is the result of an accidental degeneracy in the single particle electronic structure, see Supplementary Note 3. The unpaired electrons occupy two defect states that belong to two different irreducible representations and are localized on distinct first neighbour atoms of the defect, see the spin density in Fig.~\ref{fig:fig3}a. As a result, the exchange energy is marginal in the ground state and gives rise only to a 3.3~meV splitting between the ground state triplet and the corresponding lowest energy singlet state. The wave-functions of the low energy optical excited states exhibit similar characteristics that explain the spectrum depicted in Fig.~\ref{fig:fig2}a.  

Since the triplet and singlet states in each manifold have nearly the same spatial wave-functions and atomic configurations,  spin-orbit coupling and phonon mediated inter-system crossing (ISC) between these states are forbidden in leading order (El-Sayed's rule). Consequently, the lowest lying S0 singlet excited state close to the triplet ground state T0 is long-lived.  Indeed, we find the lifetime of the S0 state to be as long as $39.5$~ms at low temperature, which remains in the same order of magnitude at room temperature. This indicates that the S0 singlet state can serve as a classical memory to store population information.

By analysing the potential decay pathways from the excited states, we find that the states in the lowest lying ES1 manifold can directly decay non-radiatively to the ground state and its lifetime is expected to be in the pico-second range, see Supplementary Note 4. The fast decay suppresses any optical emission from the T1 triplet and S1 singlet states. On the other hand, the ES2 manifold is optically active and gives rise to zero-phonon luminescence (ZPL) lines from the T2 triplet and the S2 singlet states at $1.578$~eV and $1.586$~eV, respectively. The high temperature PL spectrum of the triplet emission is depicted in Fig.~\ref{fig:fig2}b, which shows a $60$~nm red-shifted PL phonon side-band compared to the spectrum of the boron vacancy center. The ES3 manifold decays dominantly to the ES2 manifold via non-radiative processes.

Fig.~\ref{fig:fig2}c depicts the most relevant radiative and non-radiative decay pathways of the VB2 defect. Besides the optical decay the system can decay non-radiatively from the ES2 manifold through the ES1 manifold to the ground state manifold. This process reduces the quantum efficiency of the centre for which we obtain $0.15$ and $0.12$ for the triplet and the singlet transitions, respectively.  Furthermore, since the ISC rates between the singlet and triplet channels are at least three orders of magnitude smaller than the spin-conserving decay rates, see Fig.~\ref{fig:fig2}c, the singlet and triplet emissions are decoupled from each other to a large degree. This effect and the differences in the optical lifetimes and the emission energies of the T2 triplet and the S2 singlet states can be used to realize single-shot read-out of the singlet-triplet occupation, see Supplementary Note 4 for more details.  

Since the electronic structure of the VB2 configuration does not exhibit dark long-lived metastable states the photon-count observed in experiment is limited by the radiative lifetime, two photon excitation processes, and collection efficiency of the emitted photons in saturation. Due to the high photon collection efficiency achievable in hBN~\cite{vogl_room_2017,li_near-unity_2019} and the $\approx$3~MHz radiative rate of the VB2 centre, the saturated photon-count of a single VB2 center is expected to be in the order of $100$~kcps. 

The ISC transitions between the singlet and triplet states are found to be strictly selective on the spin sub-levels of the triplet states, i.e.\ only the $\left| m_{\text{S}} = 0 \right\rangle$ state is connected with the singlet channel.  Consequently,  continuous optical excitation of the VB2 center leads to spin-polarization in the triplet ground state. Furthermore, since the triplet and singlet radiative relaxation rates are slightly different in the ES2 manifold, see Fig.~\ref{fig:fig2}c, the luminescence of the VB2 defect depends on the spin state. The saturated spin read-out contrast is found to be 4.6\%, see Supplementary Note 4. Since the contrast is set by the relative luminescence intensity of the singlet and triplet manifolds, for which we do not expect considerable temperature dependence, and not by the relative rates of various non-radiative decay processes, the observed 4.6\% contrast may be preserved even at room temperature. Finally, we note that fast ISC transitions and fast optical spin initialisation may be achieved by pumping though the ES3 manifold, where ISC rates are found in the MHz range. Such excitation, on the other hand may not be advantageous for read-out due to the enhanced non-radiative decay rates that may substantially reduces the quantum efficiency of the defect.

\subsection{VB2 centre as a qubit}
\label{sec:qubit}

\begin{figure}[h!]
\begin{center}
\includegraphics[width=1.0\columnwidth]{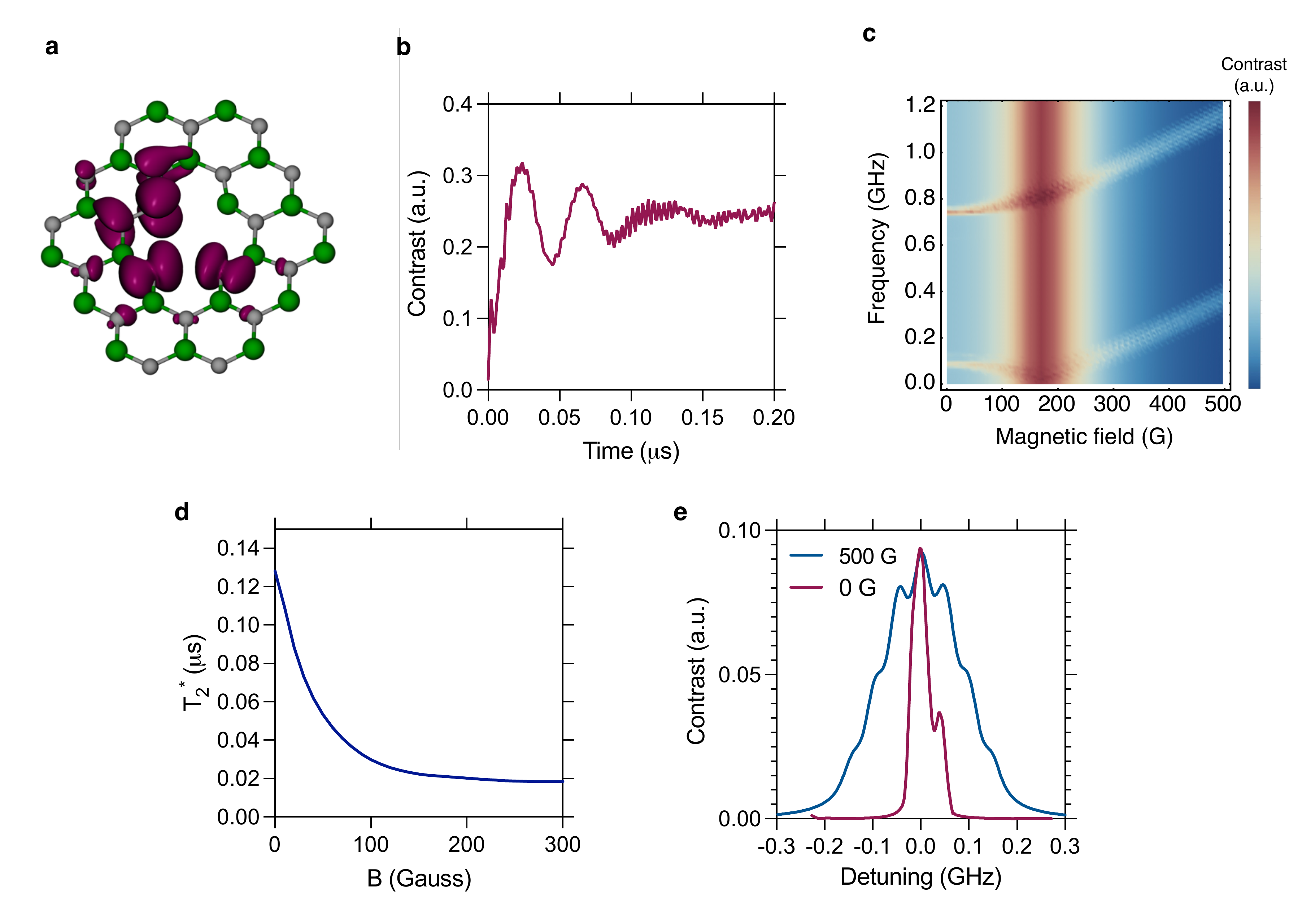}
\caption{ {\bf Qubit properties of the VB2 centre.  } {\bf a},  Spin density of the triplet ground state of VB2 centre.  {\bf b},  Rabi oscillation of the electron spin at zero magnetic field.  {\bf c} Magnetic resonance map of the VB2 centre as a function of the applied magnetic field (horizontal axis) and the microwave frequency (vertical axis).  The system is driven by an in-plane polarized microwave field. The broad frequency independent signal at $160$~G is due to the hyperfine mixing of the spin states observed at the level anti-crossing of the electron spin levels. {\bf d}, Magnetic field dependence of the free induction decay time ($T_2^{\star}$). The coherence time enhances by a factor of $7.5$ close to $B=0$ due to the transverse zero-field splitting that partially suppresses the hyperfine coupling. {\bf e}, Magnetic resonance signal at $0$~G and $500$~G magnetic field values. The width of the zero-field electron spin resonance signal is 23.8\% of the width of the signal at $500$~G. This effect gives rise to a $4.2$ times enhancement of the sensitivity at zero magnetic field.  
\label{fig:fig3}  }
\end{center}
\end{figure}

As we have shown the spin states of the ground state triplet of the VB2 centre can be initialised and read-out optically; therefore the spin states of the defect complex implement an optically addressable qubit in hBN.  In the following, we study the spin and qubit properties of the centre. Fig.~\ref{fig:fig3}a depicts the spin density of the triplet ground state that clearly shows low symmetric characteristics, in contrast to the existing point defect qubits that exhibit (quasi-)three-fold rotation symmetry. The spin Hamiltonian of the triplet ground state is discussed in Supplementary Note 5 in details.

The low symmetry configuration has a significant effect on the zero-field splitting (ZFS) interaction whose preferential quantization axis is in-plane and the corresponding conventional ZFS parameters are $D = 1.394$~GHz and $E = 78.2$~MHz.  Note, however, that the preferential quantization axis of the spin-orbit interaction is perpendicular to the plane. Hereinafter, we use this latter quantization axis, in which the electron spin is polarized in the $\left| m_{\text{S}} = 0 \right\rangle = \left|  0 \right\rangle $ spin state.  In this basis a new set of non-conventional ZFS parameters can be defined as $\widetilde{D} = -814$~MHz and $\widetilde{E} = -658$~MHz that provides the same level structure as the conventional ZFS parameters in the corresponding quantization basis.  

The large effective transverse ZFS $\widetilde{E}$ mixes the $\left| m_{\text{S}} = \pm 1 \right\rangle = \left| \pm 1 \right\rangle$ spin states and gives rise to new quantum states $\left| \pm \right\rangle = \left| +1 \right\rangle \pm \left| -1 \right\rangle$.  By applying a microwave field that is polarized in perpendicular to plane direction, spin transitions can be resonantly driven between the $ \left|  0 \right\rangle $ and $ \left|  \pm \right\rangle $ state at zero magnetic field. Rabi oscillation can be observed in the optical signal due to the microwave drive, see Fig.~\ref{fig:fig3}b. In Fig.~\ref{fig:fig3}c, we depict the magnetic field dependence of the resonance signal. As the magnetic field enhances, the zero-field mixing of the $m_{\text{S}} = \pm 1$ states reduces, the states become the eigenstates of the Zeeman Hamiltonian term, and eventually depend linearly on the magnetic field.  On the other hand, as can be seen in Fig.~\ref{fig:fig3}c, the spin states are independent of the magnetic field in first order at $B \approx 0$. The zero first-order Zeeman (ZEFOZ) transitions arising at this field are protected from magnetic fluctuations that results in an elongated coherence time of the qubit states. We study this effect through the magnetic field dependence of the free induction decay time ($T_2^{\star}$) in Fig.~\ref{fig:fig3}d.  As can be seen, the coherence time enhances by a factor of $7.5$ at $B=0$ and reaches as high as $128$~ns. Note that the coherence time may be further elongated by driving a transition between $ \left|  + \right\rangle $ and $ \left|  - \right\rangle $ states with a parallel to plane polarized microwave field. These coherence protection schemes are indispensable in hBN as it possesses nuclear spins in its lattice with 100\% natural abundance.  We note furthermore that the electron spin resonance linewidth is reduced by a factor of $4.2$ at zero magnetic field, see Fig.~\ref{fig:fig3}e.  

Note, furthermore, that the magnetic resonance map in Fig,~\ref{fig:fig3}c exhibits a broad driving frequency independent signal at $B \approx 160$~G. This signal is due to hyperfine coupling of the environment whose effect is enhanced at the anti-crossing of the ground state spin levels observed at $B \approx 160$~G, see Supplementary Note 5. The interplay of the optical pumping and the strong hyperfine coupling at this magnetic field may give rise to efficient polarization of surrounding spin bath. Low magnetic field hyperpolarization of a hBN sample can be utilized in sensitivity enhanced NMR and MRI measurements~\cite{bucher_hyperpolarization-enhanced_2020}.

\subsection{VB2 qubit as a sensor} Finally, we study the sensitivity of the VB2 centre. To this end, we first determine the pressure and electric field dependence of the ZFS parameters, for which we obtain $d \widetilde{D} / dp = 9.98$~MHz/kBar ($-8.61$~MHz/kBar) and $d \widetilde{E} / dp = 8.50$~MHz/kBar  ($-9.36$~MHz/kBar) for pressure applied parallel (perpendicular) to the V$_{\text{B}}$-V$_{\text{N}}$ axis and $d\widetilde{D} / d \mathcal{E} = 0.04$~Hz~cm/V  ($-3.45$~Hz~cm/V) and $d \widetilde{E} / d\mathcal{E} = 3.81$~Hz~cm/V  ($7.52$~Hz~cm/V) for static electric field $\mathcal{E}$ set parallel (perpendicular) to the  V$_{\text{B}}$-V$_{\text{N}}$ axis. The stress coupling parameters of the VB2 centre and the V$_{\text{B}}$ centre~\cite{gottscholl_spin_2021} are comparable,  however, these defects couple two orders of magnitude stronger to the stress field than the NV center in diamond~\cite{UdvarhelyiPRB2018}. The electric field coupling parameter of the VB2 centre along special directions is comparable to the NV centre's electric field coupling strength, however, in general it is less than an order of magnitude smaller.

The shot noise limited sensitivity of a single VB2 sensor in continuous wave magnetic resonance measurement scheme can be obtain according to the formula~\cite{dreau_avoiding_2011,pham_magnetic_2013}
\begin{equation}
\eta_{\text{MR}} =  \left(\frac{d\nu_{0+}}{d\mathcal{P}}\right)^{-1 }\frac{\Delta\nu}{  C \sqrt{\mathcal{R}}} \text{,}
\label{eq:sens}
\end{equation}
where $\nu_{0+}$ is the resonance frequency, $\mathcal{P}$ is the perturbation to be measured, $\Delta\nu$ is the full width at half maximum of the magnetic resonance curve, which takes $49.5$~MHz at zero magnetic field for electric field, pressure, and temperature sensing and $205$~MHz at high magnetic field values for DC magnetic field sensing, $C \approx 0.05$ is the spin read-out contrast,  and $\mathcal{R}$ is photon-count that can reach $100$~kcps in saturation. Using this formula and the values specified above, for pressure, DC electric field, and DC magnetic field sensitivity we obtain $30$~MPa/Hz$^{1/2}$,  $907$~kV/cm/Hz$^{1/2}$, and $463$~$\mu$T/Hz$^{1/2}$, respectively,  for a single VB2 defect in hBN in continuous wave sensing mode.  

\section*{Discussion}

The sensitivity of the VB2 centre in hBN in continuous wave measurement scheme is lower than that of the NV centre in diamond, dominantly due to the enhanced magnetic resonance linewidth, typical in hBN, and the slightly lower photon-count of the VB2 centre. On the other hand, it may still become the leading contender for high spatial resolution sensing in near surface and low dimension applications. The dipole allowed optical transition, the relatively short radiative lifetime, the saturated high-spin read-out contrast, and the narrowed-down magnetic resonance signal at zero magnetic field are all in favour of the VB2 centre. Indeed, the reported $3$~MHz radiative recombination rate is $33$ times faster~\cite{Reimers2020}, the 4.6\% high temperature read-out contrast is $30$ times larger~\cite{gottscholl_spin_2021,liu_temperature-dependent_2021}, and the magnetic resonance linewidth is about $4$ times narrower~\cite{gottscholl_initialization_2020} than that of the V$_{\text{B}}$ centre in hBN. Combining all these factors in Eq.~(\ref{eq:sens}), the saturated high temperature single defect sensitivity of the VB2 centre may exceed two orders of magnitude enhancement compared to the leading contender in hBN.

Pulsed sensing protocols and decoherence protection techniques can be utilized to further enhance the sensitivity of the VB2 centre. The low-symmetry structure offers new functionalities that have not been reported in two dimensional wide-band gap semiconductors to date. The long-lived bright singlet manifold can be utilized as a memory and may lead to the storage of spin state population information beyond the spin relaxation time of the ground state triplet. This may be utilized in single-shot read-out of the spin states when the centre is coupled with an optical cavity. Furthermore,  dressing of the $\left| \pm \right\rangle$ states with resonant microwave drive may be used in filtering out low frequency magnetic field fluctuations~\cite{miao_universal_2020} and gives rise to a highly protected subspace in hBN.

Fabrication of point defect qubits in 2D materials offers new possibilities in nanometer scale sensing.  For example,  few-layer hBN samples with pre-fabricated sensors may be moved on top of nano-scale structures or installed in van der Waals heterostructure in order to obtain high spatial resolution information not accessible with other means.  The VB2 centre, already fabricated in single layer hBN, with superior qubit properties provides a suitable sensor for such applications.

\section*{Summary}
\label{sec:disc}

In summary, we investigated the peculiar properties of the VB2 centre directly observed in our STEM measurements in a free-standing single-sheet hexagonal boron nitride. We showed that the triplet ground state of the VB2 defect implements an optically addressable spin qubit with novel characteristics. The possibility of efficient coherence protection mechanism and dipole allowed optical transition makes VB2 configuration the most suitable candidate today for emerging point defect qubit applications such as sub-nanoscale sensing in van der Waals nano and heterostructure.

\section*{Methodology}
\label{sec:meth}

In this study, we use scanning transmission electron microscopy, three state-of-the-art electronic structure computational methods, see also Supplementary Note 1, and exact spin dynamics simulations to form and determine the properties of VB2 defect.

{\noindent \bf  Sample and scanning transmission electron microscopy measurements} 
The hBN film was spin coated with PMMA, etched in ammonium borane to remove the Cu, and then transferred to a TEM grid. The PMMA was removed by acetone overnight and then the samples were annealed in vacuum overnight.  Annular dark-field scanning transmission electron microscopy (ADF-STEM) was performed using a JEOL ARM200CF STEM equipped with a CEOS probe corrector operated at an accelerating voltage of 80~kV.  ADF-STEM with a low collection angle was acquired to increase the signal from light elements and discriminate B and N atoms as darker and brighter contrast. Dwell times of $5-20$~$\mu$s and a pixel size of $0.006$~nm/px were used for imaging with a convergence semi-angle of $22.5$~mrad and a beam current of $35$~pA. Camera length of $20$~cm was used to provide large collection of low angle scattered electrons, increasing the contrast of hBN. ImageJ was used for line profile measurements and image analysis. Multislice image simulations were using QSTEM software with the accelerating voltage of $80$~kV, the convergence semiangle of $21$~mrad, the detector angle of $40-110$~mrad, the spherical aberration of $0.001$~nm, the chromatic aberration of $1$~mm, and the defocus of $-0.001$~nm. The ADF-STEM images were colourized with Gem in lookup tables using ImageJ for the better visibility.  

{ \noindent \bf  DFT calculations} We apply density functional theory (DFT) calculations to study the energetic, ground state electronic and spin properties, as well as selected excited state properties of the VB2 defect. Throughout this work, we use a plane wave basis set of $450$~eV and PAW~\cite{PAW} core potentials as implemented in VASP~\cite{VASP} and HSE06 hybrid exchange-correlation functional~\cite{HSE03} with $0.32$ exact exchange fraction~\cite{WestonPhysRevB2018}. For the charge transition level calculations we use a 768 atom supercell model of bulk hBN and  the charge correction scheme by Freysoldt and colleagues~\cite{Freysoldt}. For all the other calculations, we use a 162 atom single sheet model of hBN embedding a single defect complex. In perpendicular direction, we use $30$~\AA\ supercell size. Furthermore, in single layer calculations we use DFT-D3 correction method of Grimme and colleagues~\cite{DFT-D3-Grimme}. We use a ZFS tensor calculation routine in the PAW formalism as implemented in VASP and spin contamination error correction as proposed in Ref.~[\cite{biktagirov_spin_2020}].  We note that the spin contamination error is comparable with the largest element of the ZFS tensor, thus accurate ZFS parameters can only be achieved after decontamination. The DFT values are confirmed by NEVPT2 results for the non-strained configuration.

{\noindent \bf  NEVPT2 calculations} 

The N-Electron Valence State Perturbation Theory (NEVPT2)~\cite{angeli_introduction_2001} calculations were carried out using cc-pVDZ basis set~\cite{dunning_gaussian_1989} by the ORCA program package~\cite{neese2018software}. To this end, we employed a flake model of hBN, consisting of $38$ B-, $40$ N-, and $22$ H-atoms. The underlining complete active space self-consistent field (CASSCF) calculations were performed based on the unrestricted Hartree-Fock orbitals, while we employed the CAS of $8$ electrons and $8$ orbitals and considered $6$ triplet and $6$ singlet states for the state-averaging. The spin-orbit coupling was introduced in the framework of quasi-degenerate perturbation theory, as implemented in ORCA.

{\noindent \bf  DMRG calculations} DMRG~\cite{White_1999} is a reference-free variational post-Hartree--Fock method which relies on an iterative local optimization scheme based on Schmidt decomposition~\cite{Schollwock2005}.
The method is natively suitable for describing multi-reference problems of several dozens of strongly correlated orbitals~\cite{Olivares-2015,Szalay-2015a}. 
As DMRG is capable of treating significantly larger completely active spaces compared to CASSCF-based methods, it may be used as a reference method for strongly correlated problems. 

DMRG computations are performed applying the Budapest DMRG package~\cite{budapest_qcdmrg} restricting the spin of the target states to $0$ and $1$. In the DMRG truncation~\cite{Szalay-2015a}, the density matrix is formed of the equally weighted linear combination of the targeted spin-states and the quantum information loss is preferred to be kept below threshold value $\chi=10^{-5}$.
Considering that only six defect orbitals display multi-reference character whereas the rest of the orbitals are found to be contributing to the dynamical correlation effects, a variant of the standard dynamically extended active space approach (DEAS)~\cite{legeza_Optimizing_2003} is implemented and applied to initialize environment DMRG block states. Namely, the orbitals in the DMRG chain are arranged according to their localization on the defect atoms in descending order and the DMRG system block is optimized in the configuration space  of the environment orbitals which is composed of determinants allowing up to all single orbital excitations. 
Contrary to the standard DEAS implementation, where environment block states tend to capture the essence of the multi-reference features of the corresponding orbitals, the modified initialisation scheme focuses on the description of dynamical correlations of the complete environment from the initial DMRG micro iterations.
The application of the novel DEAS protocol is favourable for the particular orbital space setups of current interest, i.e., orbitals display well distinguished static or dynamical correlation character. In practice, the relative vertical energies are found to approach convergence within the first macro iterations of DMRG owing to the effective environmental initialisation.

The input required for the DMRG calculations, i.e., Hamiltonian matrix elements, were expanded in the active space of Kohn-Sham DFT orbitals using program suite ORCA~\cite{neese2018software,Neese-2012}.
The spin-restricted Kohn-Sham orbitals expanded in cc-pVDZ basis set~\cite{dunning_gaussian_1989} were obtained assuming PBE correlation-exchange functional and fixing half-occupation on the highest-lying valence defect orbital pair.
Matrix elements were computed by MRCC program~\cite{mrcc2020,mrcc2020b} applying the complete active space protocol~\cite{Jensen2006,Barcza2021}.
In our analysis various active space setups were tested providing consistent excitation spectra, while the actual results are presented for 100 canonical orbitals filled with 188 electrons.  The NEVPT and the DMRG calculations were performed on the same flake geometry. 

{\noindent \bf  Spin dynamics simulations}  Exact spin dynamics simulations were carried out by using a spin Hamiltonian parametrized by our DFT spin coupling parameter calculations~\cite{IvadyNPJCM2018}. Magnetic resonance (MR) spectra were obtained by propagating the states of a four spin model, including the triplet electron spin and three closest $^{14}$N triplet nuclear spins, under oscillating magnetic field of various frequency and polarization direction. The MR figures show the change of the initial population of the $m_{\text{S}} = \pm 1$ states. The $T_2^{\star}$ coherence time was obtained by simulating the coherent oscillation of 178 electron-spin--nuclear-spin two-spin systems. The overall coherence function is obtained as a product of the individual coherence functions~\cite{seo_quantum_2016}. The decaying coherence function of the electron spin was fitted by a Gaussian function to obtain the  $T_2^{\star}$. No nuclear spin nuclear spin flip-flop processes are included here.

\section*{Data availability}

The data that support the findings of this study are available from the authors upon reasonable request.

\section*{Acknowledgments}

We acknowledge support from the Knut and Alice Wallenberg Foundation through WBSQD2 project (Grant No.\ 2018.0071). Support from the Swedish Government Strategic Research Area SeRC and the Swedish Government Strategic Research Area in Materials Science on Functional Materials at Link\"oping University (Faculty Grant SFO-Mat-LiU No. 2009 00971) is gratefully acknowledged. V.I.\ acknowledges the support from the MTA Premium Postdoctoral Research Program.  A.G.\ acknowledges the Hungarian NKFIH grants No.\ KKP129866 of the National Excellence Program of Quantum-coherent materials project and the Quantum Information National Laboratory sponsored by Ministry of Innovation and Technology of Hungary as well as the support of EU Commission for the project ASTERIQS (Grant No.\ 820394).  G.B.\ acknowledges  the support from NKFIH FK-20-135496 project and from the Bolyai Research Scholarship of the Hungarian Academy of Sciences.  \"O.L.\ acknowledges financial support from the Hungarian National Research, Development and Innovation Office (NKFIH) through Grants No.\ K120569 and No.\ K134983 and the Hans Fischer Senior Fellowship programme funded by the Technical University of Munich -- Institute for Advanced Study. The development of DMRG libraries has been supported by the Center for Scalable and Predictive methods for Excitation and Correlated phenomena (SPEC), which is funded as part of the Computational Chemical Sciences Program by the U.S. Department of Energy (DOE), Office of Science, Office of Basic Energy Sciences, Division of Chemical Sciences, Geosciences, and Biosciences at Pacific Northwest National Laboratory. G.T.\ was supported by the J\'anos Bolyai Research Scholarship of the Hungarian Academy of Sciences and the \'UNKP-20-5 New National Excellence Program for Ministry Innovation and Technology from the source of the National Research, Development and Innovation Fund. The calculations were performed on resources provided by the Swedish National Infrastructure for Computing (SNIC) at the National Supercomputer Centre (NSC) and by Wigner Research Centre for Physics. We acknowledge KIF\"U for awarding us access to computational resource based in Hungary.

\section*{Author contributions}
The DFT calculations were performed by R.B., V.I.,  O.B.L., and G.T.\ with inputs from A.G.\ and I.A.A. The DFT-CAS-DMRG  and NEVPT2 calculations were performed by G.B.\ and A.P., respectively, with inputs from V.I.,  A.G., and {\"O}.L.  The STEM measurements were carried out by H.P.\ and J. H. W. The spin dynamics simulations were carried out by V.I. The results were analysed with contributions from all authors. V.I.\ wrote the manuscript with inputs from the co-authors. 

\section*{Competing interests}

The authors declare no competing interests.

\bibliography{references}

\end{document}